\documentclass[twocolumn]{aastex63}
\usepackage{graphicx}
\usepackage{amsmath}
\usepackage{apjfonts,natbib}
\usepackage{appendix}
\usepackage{amsmath}
\usepackage{amssymb}
\usepackage{bm}
\usepackage{hyperref}
\hypersetup{colorlinks,breaklinks,allcolors=blue}
\usepackage[super]{nth}

\newcommand{\hb}{H\ensuremath{\beta}}
\newcommand{\feii}{Fe\,{\footnotesize II}}

\newcommand{\lbol}{\ensuremath{L\mathrm{_{bol}}}}

\newcommand{\tsub}{\ensuremath{T_\mathrm{sub}}}

\newcommand{\rsub}{\ensuremath{R_\mathrm{sub}}}

\defcitealias{Kegerreis+2020}{K20}

\usepackage{tikz}
\usetikzlibrary{calc,fadings,decorations.pathreplacing,arrows,angles}

\tikzset{%
	>=latex, 
	inner sep=0pt,%
	outer sep=2pt,%
	mark coordinate/.style={inner sep=0pt,outer sep=0pt,minimum size=3pt,
		fill=black,circle}%
}

\shorttitle{The iron emission lines originating from the sublimated dust}
\shortauthors{He et al.}

\begin{document}

\title{An Extraordinary Response of Iron Emission to the Central Outburst in a Tidal Disruption Event Candidate}

	\correspondingauthor{Zhicheng He, Ning Jiang, Tinggui Wang}
	\email{zcho@ustc.edu.cn, jnac@ustc.edu.cn, twang@ustc.edu.cn}
	\author[0000-0003-3667-1060]{Zhicheng He}
	\affiliation{Key laboratory for Research in Galaxies and Cosmology,
Department of Astronomy, University of Science and Technology of China,
Chinese Academy of Sciences, Hefei, Anhui 230026, China}
   \affiliation{School of Astronomy and Space Sciences,
University of Science and Technology of China, Hefei, Anhui 230026, China}

\author[0000-0002-7152-3621]{Ning Jiang}
	\affiliation{Key laboratory for Research in Galaxies and Cosmology,
Department of Astronomy, University of Science and Technology of China,
Chinese Academy of Sciences, Hefei, Anhui 230026, China}
   \affiliation{School of Astronomy and Space Sciences,
University of Science and Technology of China, Hefei, Anhui 230026, China}

\author{Tinggui Wang}
	\affiliation{Key laboratory for Research in Galaxies and Cosmology,
Department of Astronomy, University of Science and Technology of China,
Chinese Academy of Sciences, Hefei, Anhui 230026, China}
   \affiliation{School of Astronomy and Space Sciences,
University of Science and Technology of China, Hefei, Anhui 230026, China}

\author{Guilin Liu}
	\affiliation{Key laboratory for Research in Galaxies and Cosmology,
Department of Astronomy, University of Science and Technology of China,
Chinese Academy of Sciences, Hefei, Anhui 230026, China}
   \affiliation{School of Astronomy and Space Sciences,
University of Science and Technology of China, Hefei, Anhui 230026, China}

\author{Mouyuan Sun}
	\affiliation{Department of Astronomy, Xiamen University, Xiamen, Fujian 361005, China}

\author{Hengxiao Guo}
     \affiliation{Department of Physics and Astronomy, 4129 Frederick Reines Hall, University of California, Irvine, CA, 92697-4575, USA}

\author{Lu Shen}
	\affiliation{Key laboratory for Research in Galaxies and Cosmology,
Department of Astronomy, University of Science and Technology of China,
Chinese Academy of Sciences, Hefei, Anhui 230026, China}
   \affiliation{School of Astronomy and Space Sciences,
University of Science and Technology of China, Hefei, Anhui 230026, China}

\author{Zhenyi Cai}
	\affiliation{Key laboratory for Research in Galaxies and Cosmology,
Department of Astronomy, University of Science and Technology of China,
Chinese Academy of Sciences, Hefei, Anhui 230026, China}
   \affiliation{School of Astronomy and Space Sciences,
University of Science and Technology of China, Hefei, Anhui 230026, China}

\author{Xinwen Shu}
\affiliation{Department of Physics, Anhui Normal University, Wuhu, Anhui, 241000, China}

\author{Zhenfeng Sheng}
	\affiliation{Key laboratory for Research in Galaxies and Cosmology,
Department of Astronomy, University of Science and Technology of China,
Chinese Academy of Sciences, Hefei, Anhui 230026, China}
   \affiliation{School of Astronomy and Space Sciences,
University of Science and Technology of China, Hefei, Anhui 230026, China}

\author{Zhixiong Liang}
	\affiliation{Key laboratory for Research in Galaxies and Cosmology,
Department of Astronomy, University of Science and Technology of China,
Chinese Academy of Sciences, Hefei, Anhui 230026, China}
   \affiliation{School of Astronomy and Space Sciences,
University of Science and Technology of China, Hefei, Anhui 230026, China}
     
\author{Youhua Xu}
	\affiliation{CAS Key Laboratory of Space Astronomy and Technology, National Astronomical Observatories, Beijing, China}



\begin{abstract}

Understanding the origin of \feii\ emission is important because it is crucial to construct the main sequence of Active Galactic Nuclei (AGNs). 
Despite several decades of observational and theoretical effort, the location of the optical iron emitting region and the mechanism 
responsible for the positive correlation between the \feii\ strength and the black hole accretion rate remain open questions as yet.
In this letter, we report the optical \feii\ response to the central outburst in PS1-10adi, a candidate tidal disruption event (TDE) taking place in an AGN at $z = 0.203$ that has aroused extensive attention. For the first time, we observe that the \feii\ response in the rising phase of its central luminosity is significantly more prominent than that in the decline phase, showing a hysteresis effect. We interpret this hysteresis effect as a consequence of the gradual sublimation of the dust grains situating at the inner surface of the torus into gas when the luminosity of the central engine increases. It is the iron element released from the sublimated dust that contributes evidently to the observed \feii\ emission. This interpretation, together with the weak response of the \hb\ emission as we observe, naturally explains the applicability of relative \feii\ strength as a tracer of the Eddington ratio. In addition, optical iron emission of this origin renders the \feii\ time lag a potential "standard candle" with cosmological implications.

\end{abstract}



\section{Introduction}

Active galactic nucleus (AGN) powered by the super-massive black hole (BH) accretion disk, is the
most luminous persistent celestial object in the universe, and can be observed up to at $z>7$ \citep{mortlock2011,banados2018,yang2020}.
Some features of AGN have the potentiality to establish as standard candles, such as the broad line region (BLR) size and luminosity relation
\citep{watson2011,czerny2013,wang2020}, nonlinear relation between UV and X-ray luminosities \citep{risaliti2019} or flux variability \citep{sun2018}.
The reliability of AGN as a standard candle depends on our understanding of AGN structure and related physical process.
Blends of \feii\ emission lines are a prominent feature in the ultraviolet (UV) and optical spectra of AGNs. 
The relative strength of optical iron, is one of the major characteristics of "Eigenvector 1" driven by the most important quantity of the 
BH accretion system, Eddington ratio \citep{boroson1992,boroson2002,shen2014}.
Despite several decades of observational and theoretical effort  \citep{boroson1992,wang1996,lawrence1997,marziani2001,boroson2002,ferland2009,shields2010,dong2011,shen2014,panda2018,panda2019}, 
the physical mechanism of \feii\ emissions has remained 
very difficult to determine. Studying the origin of \feii\ can promote our understanding of AGN structure and the related physical processes, 
hence improving the reliability of AGN as a standard candle.

Recent progresses in the time-domain surveys have led to numerous discoveries of outburst events in AGN (e.g., \citealt{kankare2017,trakhtenbrot2019}). 
Among them, tidal disruption events (TDEs) are a star occasionally ripped apart by the tidal force of SMBHs \citep{rees1988}.
TDEs in AGNs are of particular interest that offer us an unique opportunity to revisit open questions on AGN structure and related physics process in a dynamic way in months to years timescale. For example, the luminous infrared emission of AGN TDEs, originated from the dust re-processed 
emission, can yield valuable information of the dusty torus \citep{jiang2019}. 
The other fascinating characteristic associated with those TDE events is the dramatic increase of the \feii\ emission after the outburst although they will fade away later on (e.g., \citealt{drake2011, blanchard2017, kankare2017}). The transient \feii\ emission has been proposed as a natural result of sublimation of dust grains located in the inner torus due to the sharp increasing of the central emission \citep{jiang2017, jiang2019}. The iron elements primarily locked in the dust phase are released and transferred into the gas phase, contributed significantly to the \feii\ emission. However, this assumption lacks convincing evidence.

In this work, we carry out a detailed analysis on the well-known 
TDE candidate in PS1-10adi (an AGN at $z=0.203$) to study the physical process of optical \feii~emission 
and its potential cosmological applications. The paper is organized as follows.
In Section 2, we present the Data and spectral fitting. In Section 3, we analyze the emission region of optical \feii\ in TDE.
In Section 4, we present the discussions.
Throughout this work, we adopt a standard $\Lambda$CDM cosmology with with $H_{0} =70$ km~s$^{-1}$~Mpc$^{-1}$, $\Omega_{m} = 
0.3$, and $\Omega_{\Lambda} = 0.7$. 
\section{Data and spectral fitting}


\begin{figure}
\centering
\includegraphics[width=9.3cm]{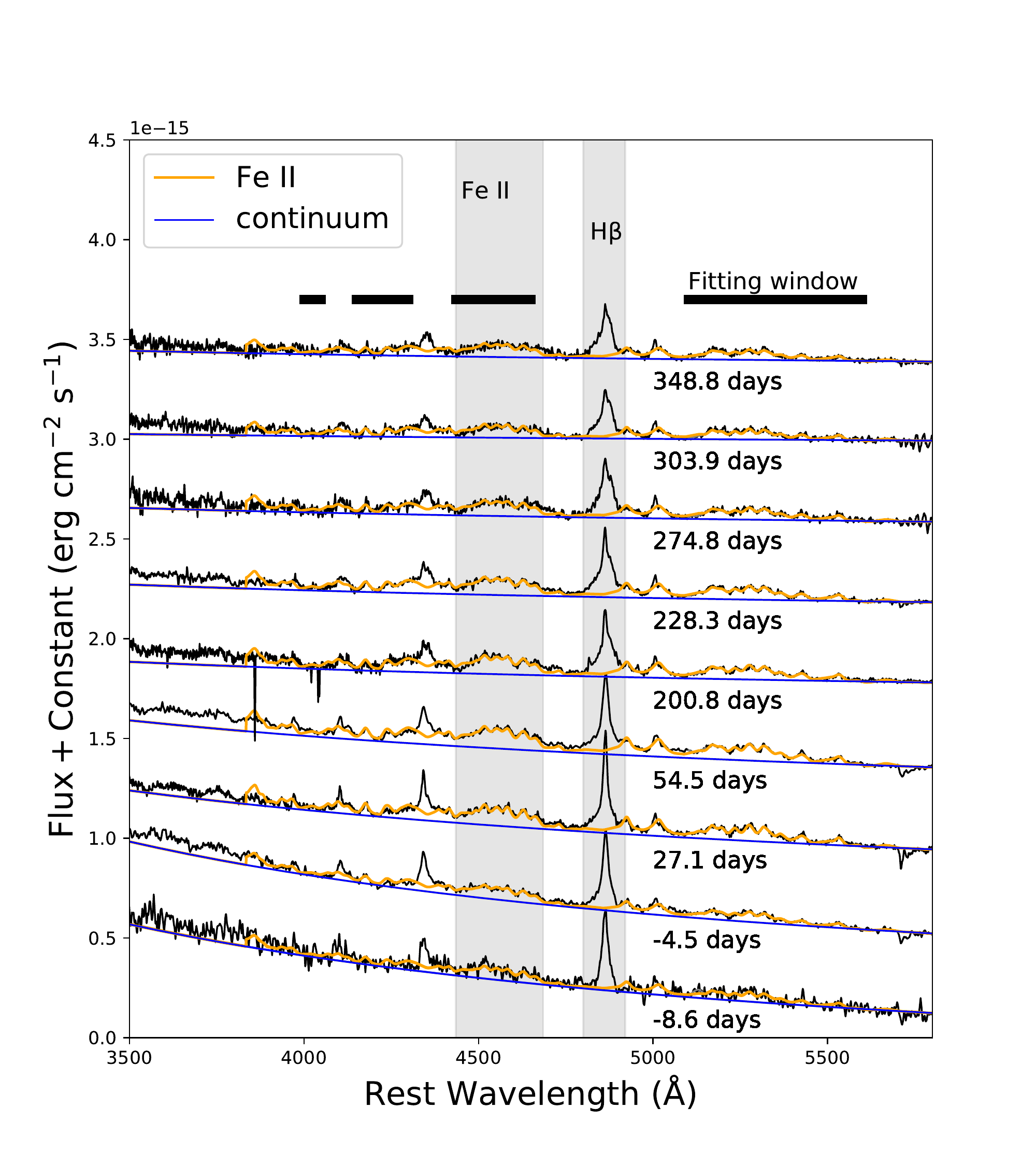}
\caption{\textbf{The spectral data of PS1-10adi (from Supplementary Fig. 2 of \citealt{kankare2017}).} 
A power-law plus the optical \feii~template \citep{boroson1992} 
to fit the continuum and \feii\ emission. The black horizontal lines mark the fitting regions which are relatively free from strong emission lines except \feii~.
The gray shadow regions mark the locations of \feii\ 4435-4685\AA\ and \hb. }
\label{fig1}
\end{figure}

PS1-10adi was initially discovered by the Panoramic Survey Telescope and Rapid Response System (Pan-STARRS) and was suggested as an energetic TDE candidate in AGN at $z=0.203$ \citep{kankare2017, jiang2019}. It stands out from AGN flares reported in the past few years because of the comprehensive observing campaign performed since its discovery, particularly its massive optical spectral data, which has even rarely covered the stage prior to the luminosity peak. The follow-up spectra taken at different stages (see Fig.~\ref{fig1}) give us an excellent dataset to explore how the \feii~strength~and relative \feii~strength, i.e., $R_{\rm \feii}=f_{\rm \feii}/f_{\rm \hb}$,
respond to the large amplitude of variation in the accretion rate. The $f_{\rm \feii}$ is calculated in the range of 4435-4685\AA\ (the left gray shadow region in Fig.~\ref{fig1}).

As shown in Fig.~\ref{fig1}, we adopt a power-law $f_\lambda \propto \lambda^{-\alpha}$ plus the optical \feii~template \citep{boroson1992} 
to fit the continuum and \feii\ emission in the wavelength regions : 
4000-4050\AA, 4150-4300\AA, 4435-4650\AA, 5100-5600\AA, which are relatively free from the strong emission lines except \feii.
We repeat this fitting process 1000 times. For each time, we add a Gaussian random error to the observed flux.
The flux of \hb\ are the observed flux by subtracting \feii\ emission and continuum: $f_{\rm \hb}=f_{\rm obs}-f_{\rm \feii}-f_{\rm con}$.
The best-fitted results and the corresponding errors are the mean and standard deviation from the 1000 fittings, respectively.
The spectral data and fitting results are shown in Table~\ref{tab1}.

\section{The emission region of optical \feii~in PS1-10adi}
\label{launch}
\subsection{The time lag of \feii\ relative to the continuum}
From Fig.~\ref{fig2}, it can be easily seen that the \feii\ emission and $R_{\rm  \feii}$ rise rapidly and reach the maximum after 55 days of the optical photometry luminosity ($L_{\rm uv-optical}$, here after $L$) peak, and then decline gradually. 
Mean while, the \hb~lemission also has a weak response to the change of central radiation. 
We notice a gap of \feii\ observations between day 55 and day 200. Thus, the real \feii\ peak might occur later than 55 days.
A theoretical calculation \citep{namekata2016} of dust sublimation radius is as follows:
\begin{eqnarray}
\begin{aligned}
\rsub  =   0.121~{\rm pc} & \left(\frac{\lbol}{10^{45}\;\mathrm{erg\;s^{-1}}}\right)^{0.5} \left(\frac{\tsub}{1800\;\mathrm{K}}\right)^{-2.8} \\
& \times \left(\frac{a}{0.1\;\micron}\right)^{-0.5}, \label{eq1}
\end{aligned}
\end{eqnarray}
where \tsub\ is the sublimation temperature of dust, $a$ is the radius of dust grain.
Given the peak luminosity $L=5\times 10^{44}\rm erg\ s^{-1}$, the theoretical torus inner radius \rsub\ is about 0.085 $\rm pc$, 
corresponding to 100 light days. \cite{kishimoto2007} found that the innermost torus radii based on dust reverberation were 
systematically smaller than the theoretical prediction of Equation \ref{eq1} by a factor $\sim$ 3. 
Thus, the observed time lag is roughly consistent with the torus inner radius.
Nevertheless, the observed time lag is not enough to prove that the \feii\ radiation is related to the torus inner radius.
In the following two subsections, we will analyze the \feii\ emission region by its evolutionary trajectory.

\subsection{The hysteresis phenomenon of the \feii\ evolutionary trajectory}
\begin{figure}
\centering
\includegraphics[width=8.5cm]{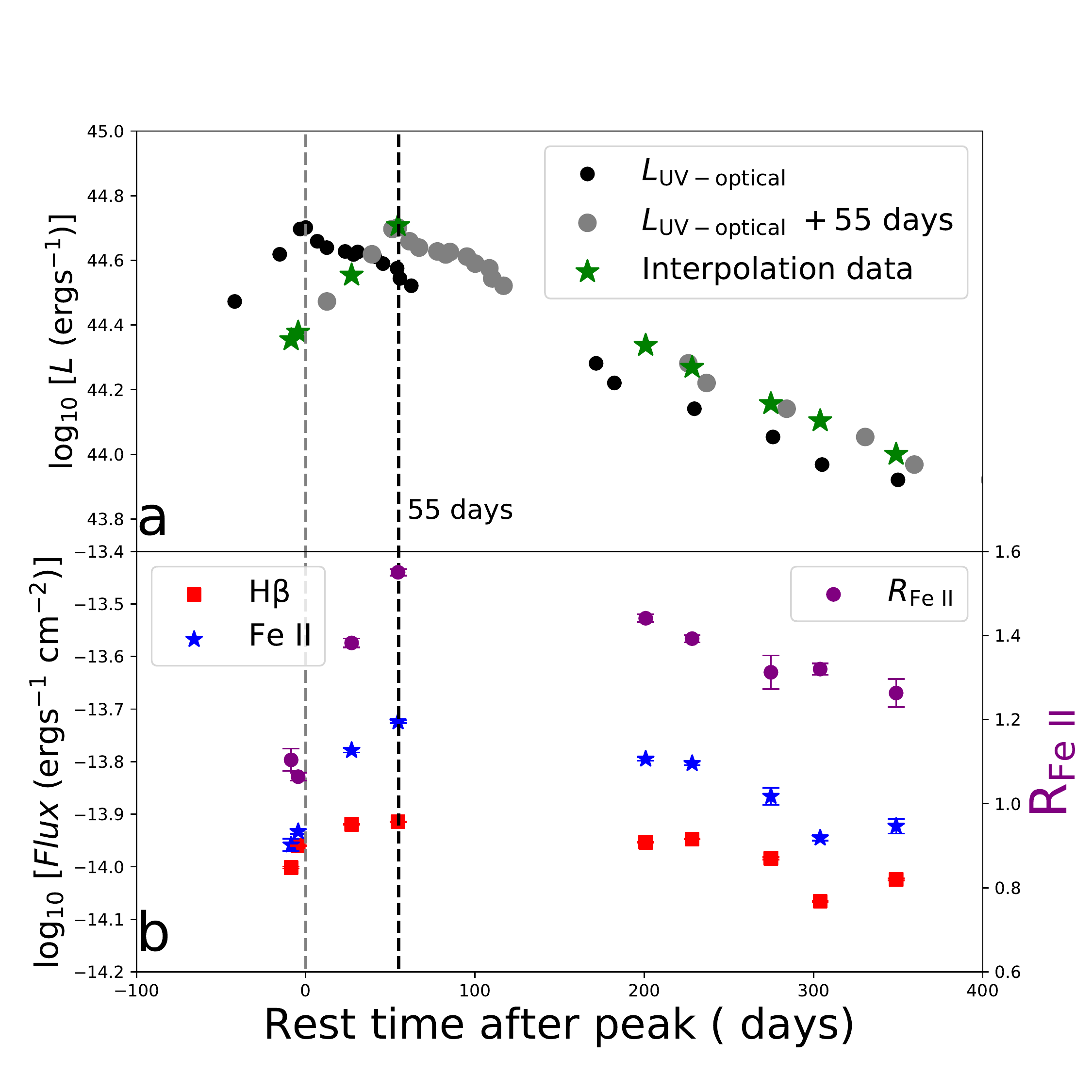}
\caption{
\textbf{The UV-optical luminosity and line flux evolution of PS1-10adi.}
Panel \textbf{a}: the black points are the UV-optical photometry luminosity. The gray points are the luminosity shifted by 55 days to 
align with the peak of \feii~emission. The green stars are the interpolation data at the corresponding time of spectral observations.
Panel \textbf{b}: the red squares, blue stars and purple points represent the \hb, \feii\ and $R_{\rm  Fe\ II}$, respectively.
The \feii\ and $R_{\rm  Fe\ II}$ rise rapidly and reach the maximum after 55 days of the luminosity peak, and then decline gradually.}
\label{fig2}
\end{figure}

As shown in Fig.~\ref{fig2}, to analyze the emission region of optical \feii, we shift the 
photometry data for 55 days to align the peaks of \feii\ emission and $L$. 
The green stars in the panel \textbf{a} of
Fig.~\ref{fig2} are the interpolation luminosities at the corresponding time of spectral observations. 

\begin{figure*}
\centering
\includegraphics[width=8.5cm]{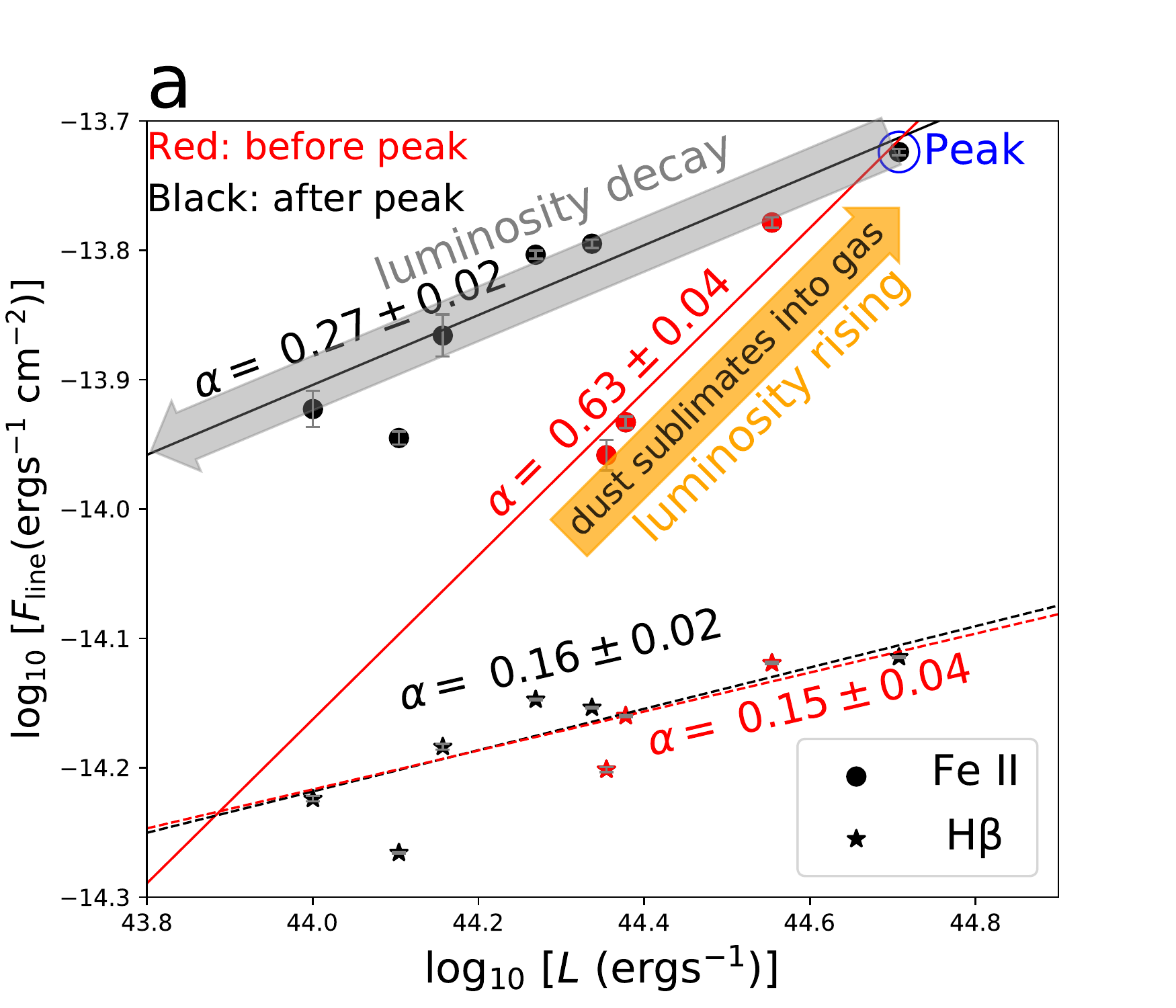}
\includegraphics[width=8.5cm]{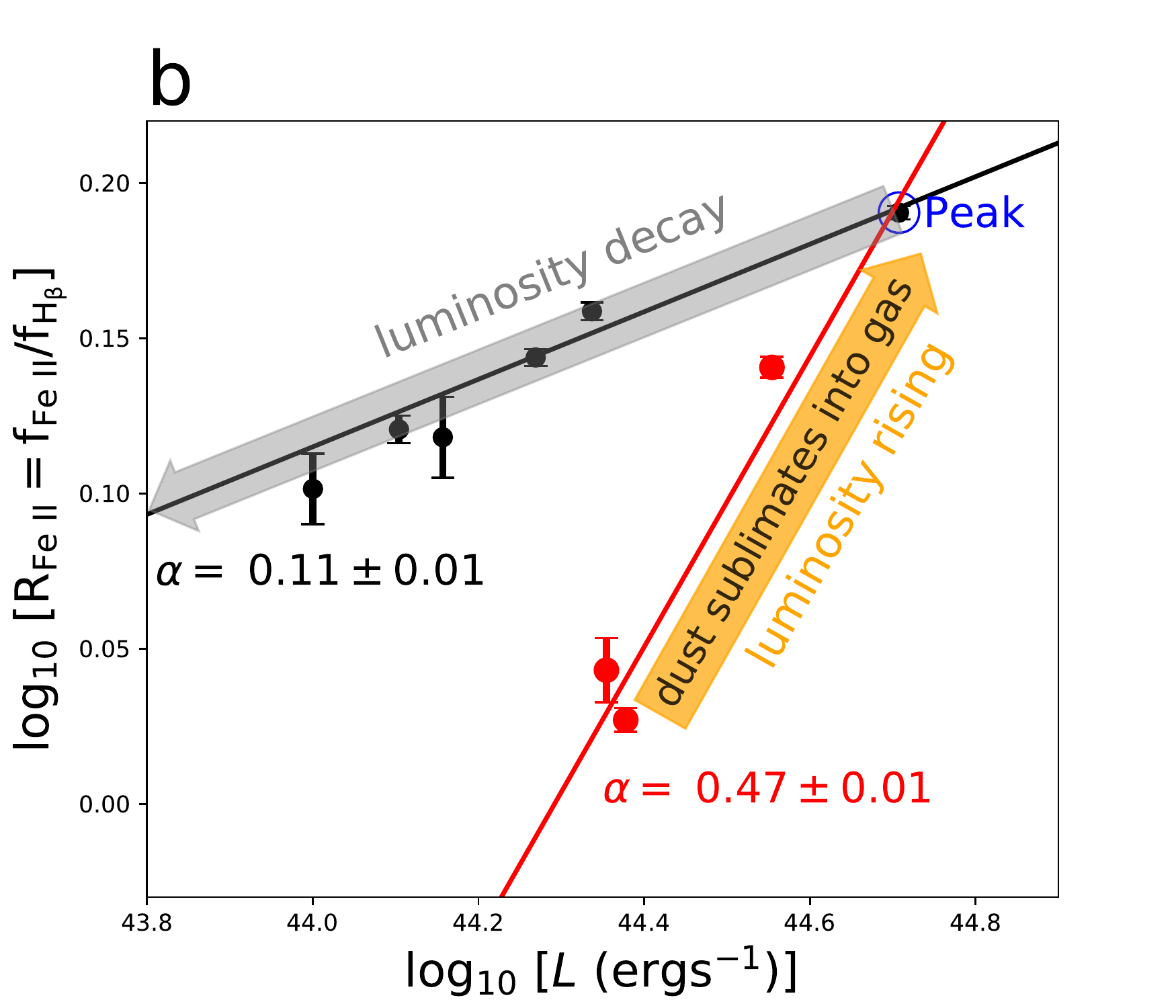}
\caption{\textbf{The hysteresis phenomenon of the \feii\ and $R_{\rm  Fe\ II}$ lines evolution in the TDE.} 
Panel \textbf{a}: the dots and stars are the \feii\ and \hb\ , respectively. The values of best-fitted $\alpha$ for each
line are marked. The $\alpha$ in the luminosity rising phase is significantly greater than that in the decline phase.
At the same luminosity, the \feii\ emission in the decline phase is significantly greater than that in the rising phase.
The evolutionary trajectory of \feii\ forms a tilted `$\Lambda$' shape (i.e., a hysteresis effect), which indicates that 
the amount of \feii-emitting gas in the decline phase is greater than that in the rising phase.
Meanwhile, there is no significant difference for \hb\ variation rate between
the two phases. This result indicates that its dominant radiation region is likely more closer than that of \feii\ and
is not dominated by the evaporated dust.
Panel \textbf{b}: similarly to \feii, the evolutionary trajectory of $R_{\rm  Fe\ II}$ also shows a hysteresis effect. 
}
\label{fig3}
\end{figure*}

We adopt the power-law function to analyze the line emission response to the change of central radiation:
$ \log_{10} f_{\rm line}=\alpha \log_{10} L+\beta $,
where $\alpha$ is the power-law slope and $\beta$ is the intercept. 
As shown in the panel \textbf{a} of Fig.~\ref{fig3}, the best-fitted $\alpha$ is $0.63\pm 0.04$ and $0.27\pm 0.02$ for \feii\ in
the luminosity rising (before peak) and decline (after peak) phases, respectively.
Interestingly, the \feii\ variation rate (described by the slope $\alpha$) in the luminosity rising phase is significantly greater 
than that in the decline phase.
At the same luminosity, the \feii\ strength in the decline phase is significantly larger than that in the rising phase.
The evolutionary trajectory of \feii\ forms a `$\Lambda$' shape (i.e., a hysteresis effect), which indicates that the amount of \feii-emitting gas
in the decline phase is larger than that in the rising phase.
Even if we shift the photometry data for 100 days (corresponding to the theoretical torus inner radius),
the hysteresis effect of the \feii\ evolutionary trajectory will still exist.

\textbf{Note that, we did not take into account the systematic uncertainties of the spectral flux calibrations in the above analysis.
As a result, the absolute flux of \feii\ or \hb\ lines maybe not so credible. However, the $R_{\rm  Fe\ II}$, i.e., relative \feii\ strength
is less affected by the systematic uncertainties of the flux calibrations.}
The evolutionary trajectory of $R_{\rm  Fe\ II}$ is shown in the panel \textbf{b} of Fig.~\ref{fig3}.
The best-fitted $\alpha$ are $0.47\pm 0.01$ and $0.11\pm 0.01$ for $R_{\rm  \feii}$ in
the central luminosity rising and decline phases, respectively. 
The `$\Lambda$' shape may be the reason of the scatter of correlation between \feii\ strength and accretion rate, especially at the low 
accretion rate state (see right panels of Fig. 10 in \citealt{dong2011}). If we divide a sample into two parts: the luminosity rising and decline phases, 
the correlation between \feii\ strength and accretion rate of the luminosity rising phase should be stronger than that of the whole sample.

\subsection{\feii\ emission dominated by the evaporated dust at inner radius of torus}

\begin{figure*}
\centering
\includegraphics[width=12cm]{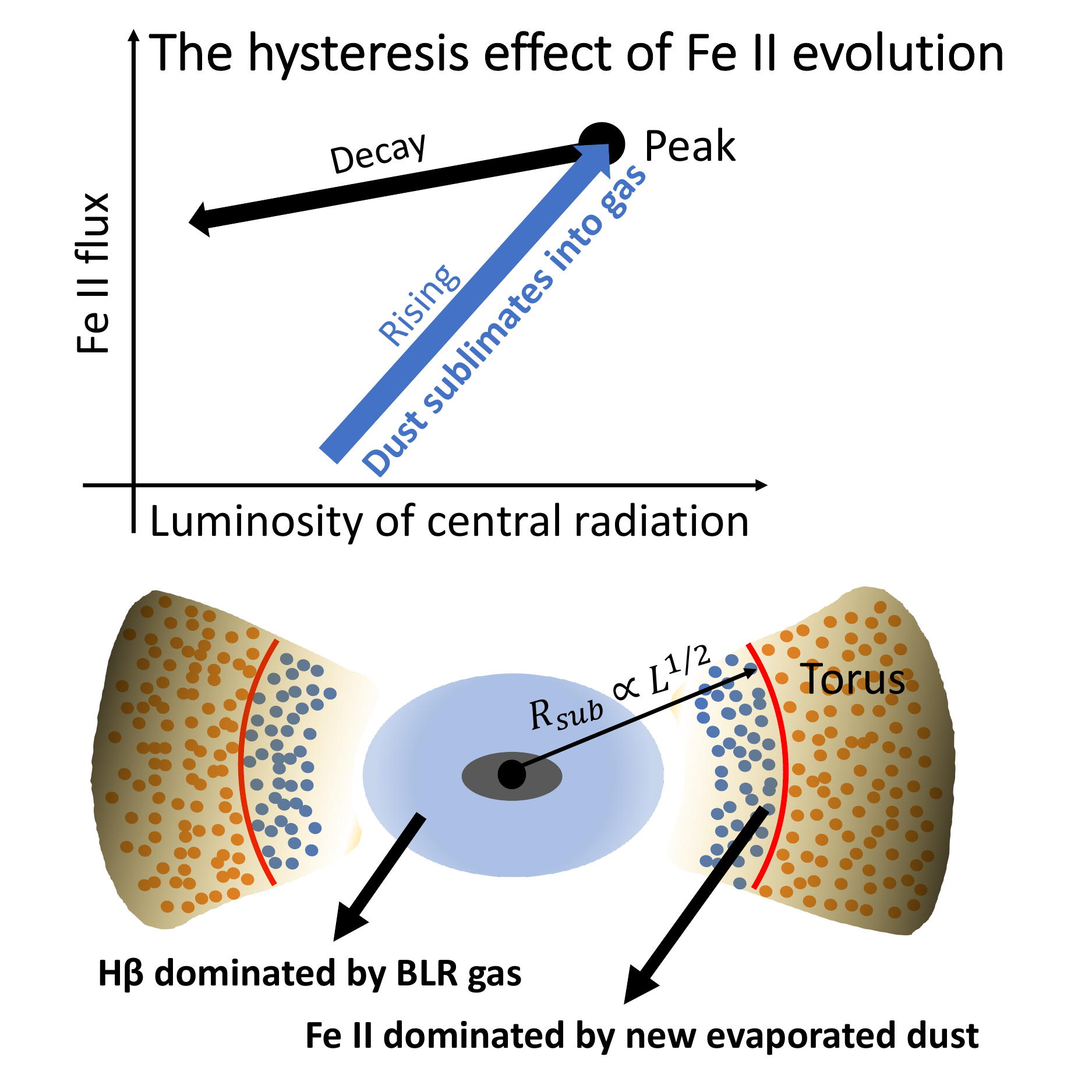}
\caption{\textbf{Schematic of \feii\ originating from the evaporated dust at inner surface of torus in TDEs.}
As the central luminosity increases, the inner boundary of the dusty torus will recede to a larger radius.
The metals released from the evaporated dust will give rise to the observed \feii\ lines.
The amount of evaporated dust reaches the maximum at the peak of central luminosity.
As a result, at the same luminosity, the \feii\ emission in the decline phase will be greater than that in the 
rising phase. The evolutionary trajectory of \feii\ forms a tilted `$\Lambda$' shape (i.e., a hysteresis effect).
Considering that the \feii~emission is dominated by the evaporated dust at inner radius of torus, which is related to the central luminosity
, i.e., $R_{\rm sub}\propto L^{1/2}$, the \feii~time lag can be adopted as a potential "standard candle" in cosmology. 
}
\label{fig4}
\end{figure*}

As shown in Fig.~\ref{fig4}, we propose a scenario to interpret the physical process 
of the \feii\ hysteresis effect. As the central luminosity increases, the dust at torus inner radius gradually 
sublimates into gas. The metals released from the evaporated dust will boost the observed \feii\ line.
The amount of evaporated dust reaches the maximum at the peak of central luminosity.
The observations of NGC4151 \citep{koshida2009,kishimoto2013} and Mrk 590 \citep{kokubo2020} suggest the
dust condensation/reformation timescale is around a few years.
Thus, the amount of evaporated dust will remain the maximum for at least a few months after the peak of central luminosity.
At the same luminosity, the \feii\ emission in the decline phase will be greater than that in the rising phase.
Meanwhile, the best-fitted $\alpha$ are $0.15\pm 0.04$ and $0.16\pm 0.02$ for \hb\ in
the luminosity rising and decline phases, respectively. There is no significant difference for \hb\ variation rate between
this two phases. This result indicates that the \hb\ emitter region is smaller than the torus inner radius and
\hb\ is not dominated by the evaporated dust, but by the BLR gas. 

The intriguing hysteresis effect of \feii~or $R_{\rm \feii}$ evolutionary trajectory strongly suggests that the 
increased \feii~emission is linked to the evaporated dust at scale of torus inner radius. In this scenario, the \feii~strength is mainly regulated by the amount of gas-phase iron (see also \citealt{shields2010}). As the central luminosity increases, the inner boundary of the dusty torus will recede to a 
larger radius. During this process, the irons released from the sublimated dust contribute evidently to the \feii~emission. 
Meanwhile, the weak response of \hb\ also implies that its dominant radiation region is likely more closer than that of \feii, which is consistent with the reverberation mapping results (e.g., \citealt{barth2013}). 
When the central luminosity increases, the response of \feii\ is stronger than that of \hb\, resulting in the increasing of 
the relative \feii\ strength $R_{\rm  Fe\ II}$.
As a result, the relative \feii\ strength has been observed as an indicator of the Eddingtion ratio of AGNs.

\begin{deluxetable*}{ccccccccccc}
\setlength{\tabcolsep}{0.03in}
\tablecaption{The spectral data and fitting result for \feii\ and \hb\ lines. \label{tab1}}
\tablehead{
\colhead{Time relative to peak (days)} & \colhead{$\log L\ (\rm erg s^{-1})$} & \colhead{  $\rm \feii\ ( \times 10^{-14}\rm erg\ cm^{-2} s^{-1})$} & \colhead{ $\rm \hb\ (  \times 10^{-14} \rm erg\ cm^{-2} s^{-1})$} & \colhead{~~~~~~~~$R_{\rm \feii}$~~~~~~~~} 
}

\startdata
    -8.6 &    44.36 &   1.101 $\pm$   0.030 &   0.997 $\pm$   0.004 &    1.10 $\pm$    0.03   \\ 
    -4.5 &    44.38 &   1.167 $\pm$   0.012 &   1.097 $\pm$   0.002 &    1.06 $\pm$    0.01   \\ 
    27.1 &    44.56 &   1.665 $\pm$   0.015 &   1.205 $\pm$   0.003 &    1.38 $\pm$    0.01   \\ 
    54.5 &    44.71 &   1.888 $\pm$   0.011 &   1.218 $\pm$   0.002 &    1.55 $\pm$    0.01   \\ 
   200.8 &    44.34 &   1.604 $\pm$   0.012 &   1.113 $\pm$   0.002 &    1.44 $\pm$    0.01   \\ 
   228.3 &    44.27 &   1.573 $\pm$   0.011 &   1.129 $\pm$   0.002 &    1.39 $\pm$    0.01   \\ 
   274.8 &    44.16 &   1.362 $\pm$   0.052 &   1.038 $\pm$   0.006 &    1.31 $\pm$    0.04   \\ 
   303.9 &    44.10 &   1.135 $\pm$   0.013 &   0.860 $\pm$   0.002 &    1.32 $\pm$    0.01   \\ 
   348.8 &    44.00 &   1.195 $\pm$   0.039 &   0.946 $\pm$   0.005 &    1.26 $\pm$    0.03  
\enddata

\tablecomments{
The luminosity corresponding to each spectrum is the result 
of interpolation in the optical photometry luminosity (green stars in Fig.~\ref{fig2}). 
}
\end{deluxetable*}

\section{Discussions}
\subsection{A Potential Cosmological Standard Candle based on \feii}

Our results suggest that the \feii~emission is boosted by the new evaporated dust at inner radius of torus in TDE.
In view of this, the time lag between the optical continuum peak and optical \feii~peak can adopted as "standard candles" in cosmology.
The AGN luminosity \lbol\ and the dust sublimation radius \rsub\ can be written as \citep{hoenig2011,honig2017}:

\begin{equation}
  \lbol=16\pi \rsub^{2}f_{\rm abs}^{-1}Q_{\rm abs};p(T_{\rm sub})\sigma_{\rm SB} T_{\rm sub}^{4},
\label{eq2}
\end{equation}
where $f_{\rm abs}$ is the fraction of incident AGN flux absorbed per dust
particle and $Q_{\rm abs};p(T_{\rm sub})$
is the normalized Planck-mean absorption efficiency of the dust.
Note that, $f_{\rm abs}$ and $Q_{\rm abs};p$ are related with both parameters approaching unity for large dust grains, 
which emit very similarly to a blackbody. Replacing \rsub\ with the corresponding time lag 
$\tau_{\rm \feii}$= $\tau_{\rm sub}=\rsub/c$, the AGN luminosity \lbol\ can be written as \citep{honig2017}:
\begin{equation}
\begin{aligned}
 \lbol =k\tau_{\rm Fe\ II}^{2}c^2,
  \end{aligned}
\label{eq3}
\end{equation}
where $k$ is a parameter absorbing $T_{\rm sub}$, $f_{\rm abs}$ and $Q_{\rm abs};p$.
A luminosity distance $D_{\rm L}$ independent of redshift is calculated as:
\begin{equation}
  D_{\rm L}=\sqrt{\frac{k}{4\pi F}}~\tau_{\rm \feii}c,
\label{eq4}
\end{equation}
where $F$ is the flux of optical continuum. The parameter $k$ can be calibrated by other cosmic distance ladder or
the spatially resolved near-IR interferometry for the inner radius of torus \citep{honig2014}.
Combination with the size of time lag and the angular diameter $\theta$ from the interferometry for the inner radius of torus, 
the angular diameter distance can be determined as: $D_{\rm A}=\tau_{\rm \feii}c/\theta$.
According to the relationship $D_{\rm L}=D_{\rm A}(1+z)^2$, the $k$ can be determined as $k=4\pi F (1+z)^4 /\theta^2 $.
The scatter of $k$ is found about 0.13 dex in the lag-luminosity relation of a sample of 17 AGN \citep{koshida2014}.
The scatter of $k$ reflects the actual object-to-object differences in hot-dust composition, geometry, and global distribution in a sample.
The low scatter value implies the tighter relation between time lag and luminosity, and the simple physics of the inner radius of torus.

\subsection{Other possible origins of the extraordinary response of iron emission}
The observed \feii\ is related not only to the central ionization continuum but also to the spatial distribution of the gas around the central
black hole accretion disk. On the one hand, the central ionization continuum may be different before and after a TDE, even at the same measured 
optical luminosity. On the other hand, the spatial distribution of gases may be different before and after a TDE.
For example, the unbound debris may fall back or the fast outflow (e.g., \citealt{hung2019}) may emerge in a TDE. 
We will need more data to test these possibilities in the future.
Anyway, the different power-laws of the \feii\ response before and after a TDE are worth more studies.

\section{conclusion}
In this letter, we present the response of \feii\ emission in the PS1-10adi TDE, an AGN at z = 0.203. 
The time lag between optical continuum peak and \feii\ or $R_{\rm \feii}$ strength  peak is consistent with the torus inner radius.
 Furthermore, we find the \feii\ variation rate in the luminosity rising phase is 
significantly greater than that in the decline phase. At the same luminosity, the \feii\ emission in the decline phase is significantly 
greater than that in the rising phase. The evolutionary trajectory of \feii\ and $R_{\rm \feii}$ show an intriguing hysteresis effect.
This result strongly suggests that the \feii~emission is boosted by the evaporated dust at scale of torus inner radius. Our results
reveal at least two applications of the \feii~emission in the TDEs:
\begin{itemize}
\item[1.] The indicator of Eddington ratio:
we propose that the dust sublimation of AGN torus accompanied with the central outburst plays a key role in the rapid increase
of \feii~strength. The irons, which were originally locked in the dust grains, get a chance to enter into the gas phase due to the sublimation 
and boost the \feii\ emission. The \feii\ strength is thus directly dependent on the amount of evaporated dust, 
which increases with the central luminosity. However, the evaporated region might contribute much less to the \hb\ emission. 
This indicates the \hb\ prefers significantly smaller emission radius. As a result, this scenario naturally explains the physical mechanism 
leading to the increase of \feii\ strength with Eddington ratio.
\item[2.] The potential cosmological application:
since the \feii~emission is dominated by the evaporated dust at inner radius of torus, which is related to the 
central luminosity, i.e., $R_{\rm sub}\propto L^{1/2}$, the \feii~time lag relative to the central luminosity can be
adopted as a potential "standard candle" in cosmology. 
In the future, we can follow up on the optical spectrum observation after the optical burst of a TDE. 
We are entering an age of accelerating development of time domain astronomy with advent of a batch of dedicated modern 
surveys (e.g., ZTF, LSST, WFST \citep{lou2016}). 
For instance, the predicted TDE number found by LSST every year can be a few of thousands 
(e.g., \citealt{thorp2019,bricman2020}), which could include hundreds of events in AGNs 
assuming a AGN fraction of 10\%.
Timely spectroscopic observations of them are highly encouraged to capture the peak of \feii~emission. 
The notion of \feii~"cosmological standards" can be soon tested and applied based on large sample studies. 
\end{itemize}

\acknowledgments
We thank the anonymous referee for the valuable comments and constructive suggestions.
We also thank Dr. Erkki Kankare for providing us the optical spectra data of PS1-10adi. 
Z.-C. H. is supported by NSFC-11903031 and USTC Research Funds of the Double First-Class Initiative YD 3440002001.
N. J. is supported by NSFC-12073025. T.-G. W. is supported by NSFC-11833007. 
G.-L. L. acknowledges the grant from the National Natural Science Foundation of China (No. 11673020 and No. 11421303) and 
the Ministry of Science and Technology of China (National Key Program for Science and Technology Research and Development, 
No. 2016YFA0400700). H. -X. G. acknowledges the NSF grant AST-1907290. 
M. -Y. S. is supported by NSFC-11973002. X. -W. S is supported by NSFC-11822301. Y.-H. X is partially supported by the NSFC-U1731127.

\clearpage
\bibliography{feii}{}
\bibliographystyle{aasjournal}

\end{document}